\documentclass[12pt]{article}
\usepackage{a4wide}
\usepackage{amssymb}
\usepackage{graphicx}
\begin{document}

\newcommand{\lpl}{\ell_{\rm P}}
\newcommand{\md}{{\rm d}}

{\renewcommand{\thefootnote}{\fnsymbol{footnote}}
\medskip
\begin{center}
{\LARGE  A Momentous Arrow of Time\footnote{Chapter contributed to ``The Arrow of Time'' Ed.\ L.~Mersini-Houghton and R.~Vaas (Springer-Verlag)}}\\
\vspace{1.5em}
Martin Bojowald\footnote{e-mail address: {\tt bojowald@gravity.psu.edu}}
\\
\vspace{0.5em}
Institute for Gravitation and the Cosmos,\\
The Pennsylvania State
University,\\
104 Davey Lab, University Park, PA 16802, USA\\
\vspace{1.5em}
\end{center}
}

\setcounter{footnote}{0}

\begin{abstract}
  Quantum cosmology offers a unique stage to address questions of time
  related to its underlying (and perhaps truly quantum dynamical)
  meaning as well as its origin. Some of these issues can be analyzed
  with a general scheme of quantum cosmology, others are best seen in
  loop quantum cosmology. The latter's status is still incomplete, and
  so no full scenario has yet emerged. Nevertheless, using properties
  that have a potential of pervading more complicated and realistic
  models, a vague picture shall be sketched here. It suggests the
  possibility of deriving a beginning within a beginningless theory,
  by applying cosmic forgetfulness to an early history of the
  universe.
\end{abstract}
\section{Introduction}

Time in quantum theory is something of a black sheep. In
non-relativistic quantum mechanics it remains a classical parameter
labelling the evolution of states, but is not allowed to fluctuate as
position does. Even in relativistic systems and quantum field theory,
time often appears as a disciplined parameter trained to order events,
much as it is used in classical physics. Crucial for particle physics
is the direction time provides for interaction events scattering
initial states into final ones. But any directedness is simply put
into the formalism. At the level of elementary reactions, time knows
no order: if a reversal of time were allowed, events would still occur
in any way, nearly unchanged.\footnote{The laws are, of course, not
completely time reflection symmetric, which might be exploited in the
context discussed here \cite{CPArrow}.} It is only our choice of
initial and final states which determines a scattering amplitude. 

All this is different on the macroscopic level and especially in
cosmology. Here, structures change with a trend. One often thinks of a
simple initial state evolving into complexity, a puzzle to be
explained by an arrow of time. If this is to be derived rather than
postulated, a theory of initial states is required.

The main part of this contribution will be an exploration of the
possibility that true quantum degrees of freedom, those such as
fluctuations which completely lack a classical analog, could play a
role of or for time.  In this way, we will take seriously quantum
space-time, not (just) as a new and possibly discrete structure but as
a fully dynamical quantum entity. More specifically, cosmological
models will lead us to an analysis of quantum correlations as
quantities changing with a trend.  If consistently realized, such a
perspective is very different from the traditional ones regarding
time: time would be inherently quantum; it would not exist in a
classical world. In semiclassical physics, it remains only as a shadow
of the quantum physics that lies beneath.

We will take advantage of a useful description of quantum dynamics
(sketched in the Appendix) based on the evolution of characteristic
quantum variables, rather than whole but partially redundant wave
functions. The same kind of description can be used to explore the
nature of non-singular big bangs. Such events, while still playing the
role of the moment of commencement for the part of the universe
accessible to us, can no longer be viewed as entire initial states of
the universe: with the singularity being resolved, there is a universe
before the big bang.  But specific realizations of such scenarios do
have derived features of special initial states as they may be posed
at the big bang. In this way, dynamical properties give insights into
the question of initial states and the directed evolution that ensues.
Especially the phenomenon of cosmic forgetfulness shows that much of
the state before the big bang remains hidden after the big bang.
Without remembrance, the arrow of time might well be considered blunt
--- or do we just see the blunt end of a reversed arrow?

By its nature, our analysis will be incomplete and preliminary. No
clear scenario emerges yet; just several indications exist. But they
may show that the topics touched here are still worth pursuing.

\section{The problem and the arrow of time}

Many questions are to be addressed in the context of time. The most
important one is, of course, the aptly named problem of time
\cite{KucharTime}. It arises mainly in canonical formulations of
gravity and attempted quantizations, but its nature reaches
farther. Independently of technical aspects, it is about the question
whether there is an unambiguous degree of freedom in generally
relativistic theories which can play the role of time, or of a
parameter whose values arrange causally related events and thus separate
the past, present and future.

This is to be distinguished from the question of the arrow of time
\cite{OpenArrow}, which irreversibly orders events already separated
into past, present and future by the time variable. Such an arrow is
often related to thermodynamical questions via entropy, or to the
selection of special initial states in quantum cosmology. The question
of the arrow of time builds on an existing time variable and is thus
to be separated from the problem of time, which is more basic. In this
contribution, we start with a discussion of the problem of time.

The problem initially arose in canonical quantizations of gravity,
with a dynamics governed only by a Hamiltonian constraint, not by a
true Hamiltonian.  Thus, quantum states do not seem to evolve,
quite obviously in conflict with the perception of
change.\footnote{It is interesting to note that the problem of time
  and motion becomes pressing when quantum gravity is considered.
  Quantum gravity is often tied to another expectation, that of
  discreteness or an atomic nature of space-time. Maybe solving the
  problem of time would lead us to establishing an atomic nature of
  time? If so, this would be reminiscent of a much older debate among
  pre-sokratic philosophers: Parmenides denied any reality to motion
  and change, which he logically argued to be pure illusion. His most
  basic statement was that nothingness does not exist, and so a body
  cannot move from where it is now to a place of empty space which was
  thought not to exist. The logical conflict was resolved by the
  atomists who accepted the notion of empty space and were led to the
  concept of material atoms.} Without any notion of space-time
coordinates in canonical quantizations of gravity, which provide
operators for geometrical quantities derived from the space-time
metric but nothing for coordinates, the usual way out by
coordinatizing time is blocked.  One is forced to identify an
appropriate time degree of freedom from the physical variables, such
as geometrical ones or matter fields. The problem is that none of them
seems to be a globally valid choice for time as an unambiguous
labelling of events.

While this problem becomes technical and pressing in canonical quantum
gravity, it is more general as well as deeper than might be indicated
at first sight. If we were able to identify a global time variable
from the physical degrees of freedom, we would be led to attributing a
new physical meaning to time. Time would cease to be a conventional
description of observed change and become a physical quantity on par
with all others. It would be subject to physical laws, and would
fluctuate in quantum theories. In that case, one might as well look
for a global time variable among the true quantum degrees of freedom
of a relativistic system, a degree of freedom such as quantum
fluctuations or correlations without a classical analog. From a
dynamical systems perspective, these are degrees of freedom in their
own right. (Such variables do play a special role from the perspective of
quantum observables since they are not obtained through expectation
values of one linear operator. But quantum fluctuations, for instance,
are certainly measurable in the same statistical sense as expectation
values.) The fundamental notion of time would crucially be tied to
quantum physics, while classical physics would have to resort to time
coordinates, a poor substitute for a truly deep notion.

Several indications exist for the squeezing of quantum matter
\cite{GasperiniEntropy,LargeSqueezing,EntropyOpen,EntropyOpenErr} or
gravitational waves \cite{GravitonEntropy} to play the role of time.
Here, following suggestions in \cite{EffCons,Recollapse} we describe
results to explore a possible relation to quantum gravity
states,\footnote{A different perspective on the importance of
  gravitatational degrees of freedom is discussed in \cite{Mersini}.}
indicating an emergent concept of time in a quantum description of
universe models. If these models and ideas are correct, the quantity
ultimately playing the role of time is not the one put in initially to
set up the evolution equations, and it is not the one used in a
classical description. This quantity, the true nature of time in the
picture proposed, does not at all exist in the classical theory. So
far, these considerations are inconclusive concerning the problem of
time. But the methods will set the stage for a discussion of the arrow
of time.

\section{Classical Dynamics}

We start with a cosmological system where the problem of time is
solved trivially: a model sourced by a free, massless scalar field
$\phi$. Thanks to the absence of any non-trivial potential, the value
of the scalar is monotonic in any time coordinate and thus can itself
be used as time. While these models are rather simple, some exactly
solvable versions provide a basis for a much more general analysis.

With a free, massless scalar as the sole matter content of an
isotropic universe with cosmological constant $\Lambda$, the expansion
history, for the different choices of spatial curvature via $k=0$ or
$k=\pm 1$, is determined by the Friedmann equation
\[
 \left(\frac{\dot{a}}{a}\right)^2+\frac{k}{a^2}= \frac{4\pi
 G}{3}\frac{p_{\phi}^2}{a^6}+\Lambda
\]
for the scale factor $a$. The dot denotes a derivative by proper time,
leading to the Hubble parameter $\dot{a}/a$. The coupling to matter is
quantified by the gravitational constant $G$, multiplying the energy
density of matter.  Here, for a free, massless scalar, only kinetic
energy is contributed via the momentum $p_{\phi}=a^3\dot{\phi}$. In
what follows, we will use $k=0$ and $\Lambda<0$ to be specific, though
not realistic.  (The case of a positive cosmological constant is very
similar to the negative sign as far as classical dynamics is
concerned, but is much more subtle at the quantum level. One can find
hints of this subtlety in the existence of different self-adjoint
extensions of the quantum Hamiltonian \cite{SelfAdFlat} or in the
dynamical behavior of quantum states.)

To employ canonical quantization later on, we now introduce the
classical canonical formulation. Choosing the (rescaled) volume
$V=a^3/4\pi G$ as configuration variable, it follows from the
Einstein--Hilbert action that its momentum is $P=\dot{a}/a$: we have
the Poisson bracket $\{V,P\}=1$. In these variables, the Friedmann
equation takes the form
\begin{equation}
 C:=(P^2+|\Lambda|)V^2-\frac{1}{12\pi G} p_{\phi}^2=0
\end{equation}
of a constraint rather than an equation of motion. A Wheeler--DeWitt
quantization \cite{QCreview} would turn this expression into an
operator $\hat{C}$ --- for instance in the volume representation where
wave functions are of the form $\psi(V,\phi)$ and $\hat{P}$ acts as
$-i\hbar \partial/\partial V$ while $p_{\phi}$ acts as
$-i\hbar\partial/\partial\phi$, with Planck's constant $\hbar$ --- and
solve the state equation $\hat{C}\psi=0$. Compared to the
Schr\"odinger equation, time is absent and change would have to be
recovered indirectly from the solution space.

What must be absent is time coordinates since they have no role in a
quantum theory of gravity, not based on classical space-time manifolds.
But other, more physical time parameters may well and should indeed
exist. Realizing this is facilitated by eliminating time coordinates
already at the classical level, and finding an alternative formulation
of classical evolution. To do so, we write equations of motion for
$V$ and $P$ with respect to the scalar $\phi$. Such equations can be
obtained by dividing equations of motion with respect to coordinate
time, such as $\md V/\md\phi= \dot{V}/\dot{\phi}$. But any reference
to coordinate times can be avoided altogether if we solve the
Friedmann equation for the momentum
\begin{equation}\label{pphi}
  p_{\phi} =\pm2\sqrt{3\pi G}\,V\sqrt{P^2+|\Lambda|} =: H(V,P) 
\end{equation}
and take $H(V,P)$ as the Hamiltonian for evolution in $\phi$. (We will
choose the $+$-sign in what follows, letting $\phi$ run along with
coordinate time.) The Hamiltonian equation of motion $\md O/\md\phi=
\{O,H\}= \partial O/\partial V\cdot\partial H/\partial P- \partial
O/\partial P\cdot\partial H/\partial V$ for any function $O$ of $V$ and $P$
then equals what we would obtain from dividing coordinate equations of
motion.

The case $\Lambda=0$ is particularly simple. It provides a quadratic
Hamiltonian $H\propto |VP|$ and thus constitudes an example of
harmonic cosmology \cite{BouncePert,BounceCohStates}. Just as the
harmonic oscillator in mechanics, it leads to an exactly solvable
quantum system --- not just in the sense that solutions can be found
in closed form, but with the much stronger property that no quantum
back-reaction occurs. The evolution of expectation values is entirely
unaffected by changing shapes of a state. In the next section we will
see what that entails for dynamics, and how perturbation theory can be
used to step from the exactly solvable model to more realistic cases
as they are obtained for $\Lambda\not=0$ or with a non-trivial matter
potential.

\section{Quantum Dynamics}

\begin{figure}
\begin{center}
\resizebox{0.8\textwidth}{!}{%
  \includegraphics{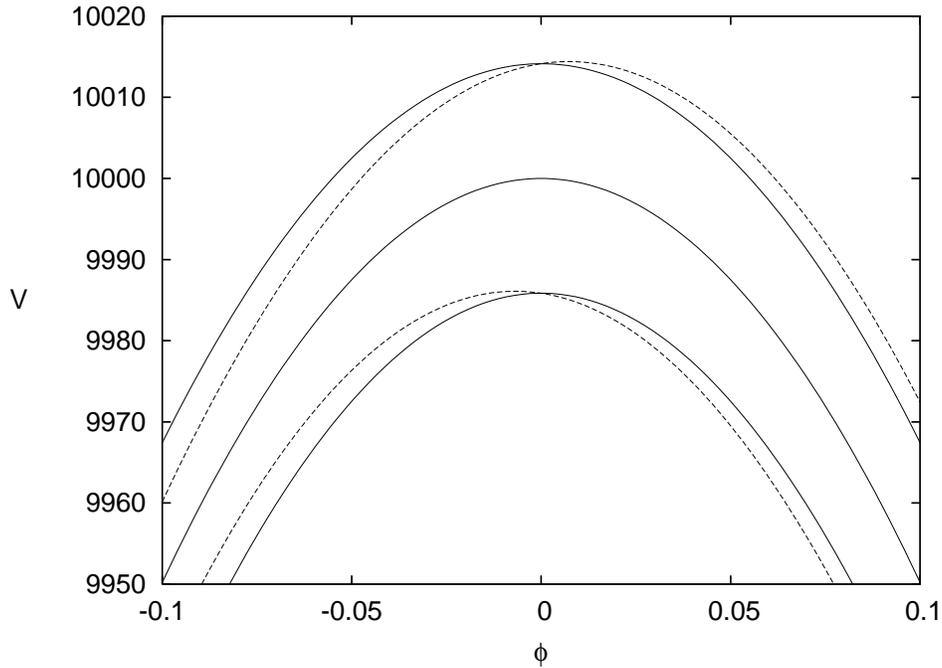}}
\caption{Dispersing wave at the recollapse of a
  Friedmann--Robertson--Walker model with $k=0$ and $\Lambda<0$. The
  top and bottom curves indicate changing spreads $\Delta V$ around the central
  expectation value trajectory of volume $V$ as a function of the
  scalar $\phi$. The latter serves as a measure for time in this free,
  massless case. For the solid lines, the state is unsqueezed, without
  quantum correlations, at the recollapse point, and fluctuations
  symmetric around the recollapse result. Non-vanishing correlations
  (dashed lines), on the other hand, lead to non-symmetric
  fluctuations.
  \label{f:Recoll}}
\end{center}
\end{figure}

We now turn to the quantum dynamics of our systems. A quantum system
is characterized by the presence of additional, non-classical degrees
of freedom which can change independently of the classical variables,
given by $V$ and $P$ above.  While the latter can be brought in
correspondence with expectation values $\langle\hat{V}\rangle$ and
$\langle\hat{P}\rangle$ in a quantum state, a whole wave function (or
density matrix) contains much more information. Indeed, while
classical phase space functions $f(V,P)$ are merely combinations of
the canonical coordinates and are fully determined if only a phase
space point is specified, products of operators in a quantum system
provide independent kinds of information. In general, for instance,
$\langle\hat{V}^2\rangle$ can take values irrespective of what the
value of $\langle\hat{V}\rangle^2$ is. The difference $(\Delta V)^2=
\langle(\hat{V}-\langle\hat{V}\rangle)^2\rangle$ is a measure for
quantum fluctuations, an important quantity in a quantum system.
Similarly, all {\em moments}
\begin{equation}\label{Moments}
G^{\underbrace{\scriptstyle V\cdots V}_a\underbrace{\scriptstyle P\cdots P}_b}=\langle(\hat{V}-\langle\hat{V}\rangle)^a
 (\hat{P}-\langle\hat{P}\rangle)^b\rangle_{\rm Weyl}
\end{equation}
defined for $a+b\geq 2$ are independent parameters of a (density)
state.  (The subscript ``Weyl'' indicates that operator products are
ordered totally symmetrically before taking the expectation value.) As
discussed in the Appendix, these moments are all dynamical, forming an
infinite-dimensional coupled system. Solutions tell us how expectation
values of a state behave, but also how the state and its moments
evolve. Fig.~\ref{f:Recoll} shows the example of a quantum
cosmological state during a recollapse, with the spreads changing
characteristically.  

From these moments we will attempt to form an
arrow of time.

\subsection{Monotonicity}

When a quantum state changes, its moments change. Just like the
expectation values, the moments must satisfy equations of motion
which, as derived in the Appendix, follow from the Hamiltonian.
Having equations of motion for the second order moments, we can check
if any one of them would serve as a good time coordinate
\cite{Recollapse}. Since quantum states tend to spread out, one may
expect fluctuations to have an interpretation of internal time. For
$G^{PP}$, however, this is clearly not the case since its change
(\ref{GPP}) depends on the sign of curvature $\langle\hat{P}\rangle$.
It would decrease in an expanding universe but increase in a
collapsing one. Around a recollapse or a bounce, this behavior cannot
be monotonic. Similarly, the sign of the rate of change of volume
fluctuations $G^{VV}$ is not unique from (\ref{GVV}) since neither
$G^{VP}$ nor $\langle\hat{P}\rangle$ is required to have a definite
sign throughout the history of a universe.

Of more interest for our purposes is the covariance, subject to
\[
  \frac{\md G^{VP}}{\md\phi} = \frac{3}{2}|\Lambda|
\frac{\langle\hat{V}\rangle}{(\langle\hat{P}\rangle^2+
 |\Lambda|)^{3/2}} G^{PP}
\]
from (\ref{GVP}).  With $G^{PP}=(\Delta P)^2$ required to be positive,
the covariance can only grow. Thus, the positivity of fluctuations, or
the uncertainty relation in even stronger form, implies a fixed
tendency for correlations.

The monotonicity of $G^{VP}$ hints at a possible role in the context
of time. In the model considered so far, it certainly does not improve
the problem of time since it can anyway be solved trivially by using
the scalar (with respect to which $G^{VP}$ now is monotonic). But if
we have a look at models with a non-trivial scalar potential
$W(\phi)\not=0$, where $\phi$ would no longer serve as global time,
one can see that $G^{VP}$ is better behaved than just $\phi$. In such
a case, with a time-dependent potential in the formulation where
$\phi$ plays the role of time, equations can be derived as before
provided that the potential is not too large
\cite{BouncePot,EffConsRel}.  The classical constraint (still for
$\Lambda<0$) now is
\[
 \left(P^2+|\Lambda|-\frac{8\pi G}{3}W(\phi)\right)V^2-
\frac{1}{12\pi G}p_{\phi}^2=0
\]
and effective equation of motion for the covariance changes to
\begin{equation}\label{GVPPot}
  \frac{\md G^{VP}}{\md\phi} = \frac{3}{2}\frac{\langle\hat{V}\rangle
(|\Lambda|-8\pi
GW(\phi)/3)}{(\langle\hat{P}\rangle^2+|\Lambda| -8\pi G W(\phi)/3)^{3/2}}
G^{PP}\,.
\end{equation}

For sufficiently small potentials, $G^{VP}$ is still monotonic for
wide ranges of evolution. Also here, this refers to monotonicity with
respect to $\phi$, which now is a good time variable only for finite
stretches between turning points in the potential.  If we approach a
turning point of $\phi$, however, the behavior changes. At a turning
point, $p_{\phi}=0$ and thus $\langle\hat{P}\rangle^2=-|\Lambda| +8\pi
G W(\phi)/3$ from the constraint. Near a turning point,
$|\Lambda|-8\pi GW(\phi)/3$, appearing in the numerator of
(\ref{GVPPot}), thus becomes negative.  Even before the turning point
of $\phi$ is reached, $G^{VP}$ according to (\ref{GVPPot}) turns
around.

Near a turning point the potential is important and there may be extra
terms in the quantum equations of motion. Our analysis at this stage
remains incomplete, but it suggests a situation as follows.  As a
global time variable through periods of oscillation of $\phi$,
$G^{VP}$ appears no better than the scalar.  But it is monotonic in a
range around the turning point and can thus be used as local internal
time, to which we may transform from $\phi$ (or other time choices)
when a turning point is approached.  Thus, it would extend its role of
time into a wider region. As a quantum variable without classical
analog, this at least suggests that time in a fully relativistic
situation can be assisted by quantum aspects.

\subsection{Before the big bang}

So far, we have discussed only low-curvature regimes where $P\ll 1$.
At larger curvature, new effects from quantum gravity and quantum
geometry are expected to take over which are not included in the
Wheeler--DeWitt quantization \cite{QCreview} understood up to now.
Loop quantum cosmology \cite{LivRev} is one such candidate for an
extension, and one of its effects is to provide higher order terms to
the Friedmann equation. Its new form then is
\begin{equation}\label{loopFriedmann}
 \frac{\sin^2(\mu P)}{\mu^2}= \frac{1}{12\pi G}\frac{p_{\phi}^2}{V^2}
\end{equation}
where $\mu$ is a length scale (see, e.g., \cite{DiscCorr}).

Such higher order terms of $P$ or $\dot{a}$ are expected in quantum
gravity if we realize the Friedmann equation as the time-time
component of Einstein's tensorial equation. Higher curvature
corrections change the action, and thus the Einstein tensor.
Correspondingly, the Friedmann equation is amended by higher order
terms. (The same reasoning would suggest higher derivative terms, too,
which generically are also present. We will, however, be dealing with
a solvable model of a free, massless scalar where they are absent
\cite{BouncePert,BounceCohStates}.) With this analogy, the expansion
parameter $\mu$ is the same as the one multiplying higher curvature
terms, and thus should indeed be dimensionfull. One may think of it as
being near the Planck length, which is in fact often assumed. But in
loop quantum gravity, it has a dynamical origin related to the
discreteness of an underlying quantum gravity state
\cite{InhomLattice,CosConst}. Generically, $\mu$ changes as the
universe expands or contracts and cannot always be close to the Planck
length. In fact, if it were, other corrections (from inverse scale
factor terms \cite{InvScale}, based on \cite{QSDV}) would have to be
considered as independent quantum corrections, which we avoid here.

The form of the higher order terms, obtained by expanding $\sin(\mu
P)$ by powers of $P$ when curvature is small, as well as the length
scale $\mu$ determining when quantum corrections become important, is
not fixed. It may be constrained further by relating such a
Hamiltonian to one that is formulated in the full theory, without any
symmetry assumptions. But this has currently not been achieved, and so
the precise form remains subject to quantization ambiguities. What we
discuss in what follows only involves generic qualitative features
which depend on some crucial aspects of the loop quantizaton but not
on the specific form. As an important effect we include lattice
refinement, leading to a possible $V$-dependence of the parameter
$\mu$: the characteristic length scale where discreteness effects
happen might depend on the volume and change dynamically
\cite{InhomLattice}.  Conceptual \cite{SchwarzN} as well as
phenomenological constraints
\cite{RefinementInflation,RefinementMatter} on the dependence exist,
and it is clear that $\mu$ cannot be $V$-independent in all models
\cite{Consistent}; but in no case has it been fixed uniquely. A
power-law dependence of $\mu\propto V^{\kappa}$ on $V$, which can
realistically describe at least bounded ranges of evolution, can be
taken into account by a canonical transformation $P\mapsto
V^{\kappa}P$, $V\mapsto V^{1-\kappa}/(1-\kappa)$ which will not change
the following results.

Now, the Hamiltonian for $\phi$-evolution is not
quadratic in $V$ and $P$ even for $k=0=\Lambda$, suggesting
non-perturbative effects at strong curvature $P\gg 1/\mu$.
Fortunately, the system is ``resummable'' \cite{BouncePert}: it is
solvable and free of quantum back-reaction if we use the variables
$V$, $J=V\exp(i\mu P)$ instead of canonical ones. These variables
satisfy a linear Poisson algebra
\begin{equation}
 \{V,J\}= i\mu J\quad,\quad \{V,\bar{J}\}= -i\mu \bar{J}\quad,\quad
 \{J,\bar{J}\}= -2i\mu V
\end{equation}
and they provide the basis for solvability even at the dynamical
level. In fact, the Hamiltonian for $\phi$-evolution, solving
(\ref{loopFriedmann}), is $p_{\phi}=2 \sqrt{3\pi G} {\rm Im}J$ which
is linear in the basic variables. Linearity implies that all these
relations have direct analogs at the quantum level:
$[\hat{V},\hat{J}]=-\mu\hbar \hat{J}$, $[\hat{V},\hat{J}^{\dagger}]=
\mu\hbar \hat{J}^{\dagger}$, and $[\hat{J},\hat{J}^{\dagger}]=
2\mu\hbar\hat{V}$ together with the Hamiltonian $\hat{H}=-i\sqrt{3\pi
G}(\hat{J}-\hat{J}^{\dagger})$.  This strong form of solvability
allows us to analyze the evolution of a state in precise terms,
especially when it approaches the classical singularity.

First, thanks to solvability there is no quantum back-reaction and
expectation value equations of motion form a closed set:
\begin{equation} \label{loopEOM}
 \frac{\md \langle\hat{V}\rangle}{\md\phi}=
   \frac{\langle[\hat{V},\hat{H}]\rangle}{i\hbar}= \sqrt{3\pi G} 
(\langle\hat{J}\rangle+\langle\hat{J}^{\dagger}\rangle)\quad,\quad
 \frac{\md \langle\hat{J}\rangle}{\md\phi}=
 \frac{\langle[\hat{J},\hat{H}]\rangle}{i\hbar} = 2\sqrt{3\pi G} 
\langle\hat{V}\rangle\,.
\end{equation}
These equations can be combined to
$\md^2\langle\hat{V}\rangle/\md\phi^2= 12\pi G \langle\hat{V}\rangle$,
easily integrating to
\[
 \langle\hat{V}\rangle(\phi)=
   \alpha\cosh(2\sqrt{3\pi G}\phi)+ \beta \sinh(2\sqrt{3\pi G}\phi)
\]
with constants of integration $\alpha$ and $\beta$ to be fixed by
initial values. Using (\ref{loopEOM}), we then obtain
\[
{\rm Re}\,\langle\hat{J}\rangle(\phi)= 
\frac{1}{2\sqrt{3\pi G}}\frac{\md V}{\md\phi}=
\alpha\sinh(2\sqrt{3\pi G}\phi)+ \beta \cosh(2\sqrt{3\pi G}\phi)\,.
\]
The imaginary part of $\langle\hat{J}\rangle$
is fixed to be
\[
 {\rm Im}\,\langle\hat{J}\rangle(\phi)= \langle\widehat{V\sin(\mu P)}\rangle=
 \frac{\mu}{2\sqrt{3\pi G}} p_{\phi}
\]
by the preserved $\phi$-Hamiltonian, using (\ref{loopFriedmann}).

The constants of integration $\alpha$ and $\beta$ determine whether or
not $\langle\hat{V}\rangle$ can reach zero, where a singularity would
occur. Due to reality conditions, these constants are not arbitrary:
classically we have $|J|^2-V^2=0$, which is to be imposed as an
operator equation $\hat{J}\hat{J}^{\dagger}-\hat{V}^2=0$ after
quantization. (Otherwise the curvature parameter $P$ would not become
self-adjoint and physical states obtained by solving the evolution
equations would not be correctly normalized.) Since the reality
condition is quadratic, it implies
\begin{equation}
0=\langle \hat{J}\hat{J}^{\dagger}-\hat{V}^2\rangle=  
\langle\hat{J}\rangle \langle\hat{J}^{\dagger}\rangle- \hat{V}^2+
  G^{J\bar{J}}-G^{VV} +\mu\hbar\langle\hat{V}\rangle
\end{equation}
with extra terms from fluctuations. (The last term arises from
ordering $\hat{J}\hat{J}^{\dagger}$ symmetrically.) A state which is
semiclassical at a given time has fluctuations of the order
$O(\hbar)$, such that the reality condition takes the classical form
up to small terms of order $\hbar$. Then, our dynamical solutions must
satisfy
\[
 ({\rm Re}\langle\hat{J}\rangle)^2+({\rm
   Im}\langle\hat{J}\rangle)^2-\langle\hat{V}\rangle^2=
   -\alpha^2+\beta^2+\frac{\mu^2}{12\pi G} p_{\phi}^2=O(\hbar)
\]
which determines $\beta$ in terms of $\alpha$. With $V_{\rm min}:= \mu
p_{\phi}/12\pi G$ and the identity $B\cosh(x+\cosh^{-1}(A/B))=
A\cosh(x)+ \sqrt{A^2-B^2} \sinh(x)$ for arbitrary $A$ and $B$, the volume is
\begin{equation} \label{Volume}
 \langle\hat{V}\rangle(\phi)= V_{\rm min}\cosh(2\sqrt{3\pi G}\phi+ \delta)
\end{equation}
with $\delta= \cosh^{-1}(\alpha/V_{\rm min})$. This function never
becomes zero, proving that the model has a bounce but no singularity.
At the bounce point, the density of the scalar field takes the value
\begin{equation}
 \rho_{\rm crit}= \frac{p_{\phi}^2}{2a^6}=
 \frac{p_{\phi}^2}{32\pi^2G^2V_{\rm min}^2}= \frac{3}{8\pi G\mu^2}
\end{equation}
which depends on the scale $\mu$ but is independent of any initial
condition. (The same behavior initially arose from numerical studies
\cite{APSII}.)

\begin{figure}
\begin{center}
\resizebox{0.7\textwidth}{!}{%
  \includegraphics{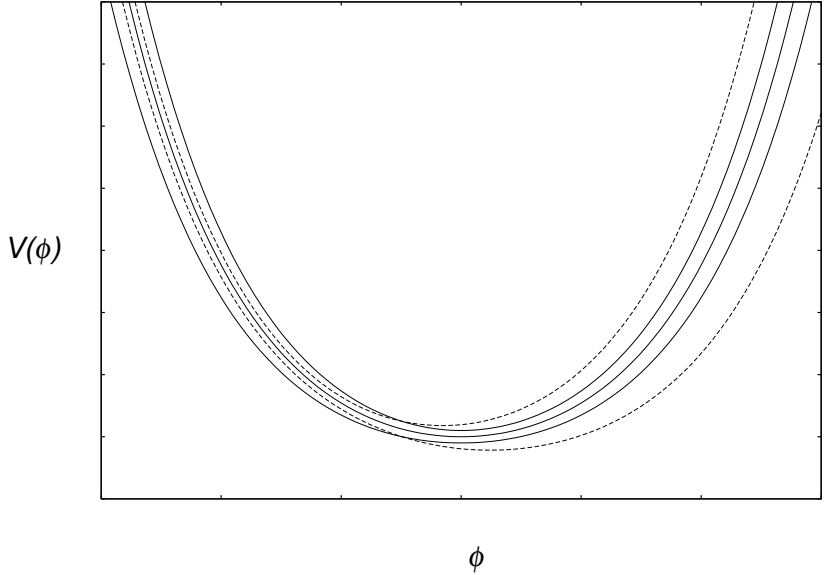}}
\caption{Dispersing through a bounce. Here, the volume $V(\phi)$ as a
  function of the scalar, again indicating time, is shown for a bounce
  rather than a recollapse as in Fig.~\ref{f:Recoll}. As before,
  the top and bottom curves indicate fluctuations $\Delta V$ around
  the expectation value $V$ --- solid curves for a state uncorrelated at
  the bounce, dashed curves for a correlated one. Fluctuations
  ``before'' the big bang may have been quite different from what they
  are ``afterwards'' --- see also Eq.~(\ref{asymm}) --- to a degree
  that can be considered forgetful.
 \label{f:EffBounce}}
\end{center}
\end{figure}

To evaluate the reality condition, we have used semiclassicality. One
might worry that this invalidates conclusions about the bounce,
typically expected to occur in a highly quantum regime. However, we
had to make assumptions about semiclassicality only at one time, which
can be arbitrarily far away from the bounce. We only need
$G^{J\bar{J}}-G^{VV}=O(\hbar)$ throughout, which is at first ensured
by an initial condition at large volume. As the state evolves, it may
become more quantum. But from equations of motion for the moments it
follows that $G^{J\bar{J}}-G^{VV}$ is a constant of motion
\cite{BounceCohStates}, even if the state spreads, making $G^{VV}$
change. Thus, if this combination is of the order $\hbar$ once, it
will remain so. In this solvable model the bounce is realized even for
states which may not be semiclassical at the bounce.

The high control in this solvable model persists at the state level.
Dispersions as well as squeezing can be followed for general states,
as well as specifically for the moments of dynamical coherent states;
see Fig.~\ref{f:EffBounce} for examples. In principle, we could thus
test how covariances evolve and whether they remain monotonic.
However, moments with easy access are now those of $V$ and $J$, not
$P$. Volume fluctuations thus can easily be studied, but the
covariance $G^{VP}$ of our earlier interest would, with $P$ related
non-linearly to $J$, be a complicated expression in terms of all the
moments involving $V$ and $J$. Nevertheless, we can find approximate
information about the behavior of covariance. Near the bounce, we have
$\mu P\sim \pi/2$ for $\sin(\mu P)$ and thus the scalar density to be
close to its maximum. This allows us to use the approximation 
\begin{eqnarray*}
 {\rm Re}\langle\hat{J}\rangle&=& \frac{1}{2}\langle\hat{V}\widehat{e^{i\mu
 P}}+ \widehat{e^{-i\mu P}}\hat{V}\rangle \sim
 \frac{1}{2}\langle e^{i\pi/2}\hat{V}i(\mu\hat{P}-\pi/2)-e^{-i\pi/2}
 i(\mu\hat{P}-\pi/2)\hat{V}\rangle\\
&=& -\frac{\mu}{2}\langle\hat{V}\hat{P}+\hat{P}\hat{V}\rangle+
 \frac{\pi}{2}\langle\hat{V}\rangle\sim -\mu G^{VP}
\end{eqnarray*}
by Taylor expansion around $\mu P\sim \pi/2$. Noting that
\[
 {\rm Re}\langle\hat{J}\rangle= V_{\rm min} \sinh(2\sqrt{3\pi G}\phi+\delta)
\]
from (\ref{loopEOM}) and (\ref{Volume}) is monotonic in $\phi$, we are
led to suggest that also the covariance on the right hand side is
monotonic. Combining all conclusions, it will thus be a good measure
for time through several cosmological phases, including recollapses
and bounces.

\section{Beyond Exactitude}

So far, we have considered a free, exactly solvable model to shine
some light on the universe at small volume. Such models rarely give
the full picture of a physical situation they may be applied to. There
are several additional ingredients to be required for a physically
reliable analysis of a whole universe through and before the big bang,
most importantly inhomogeneous configurations. No general description
of inhomogeneities is available around bounce regimes in loop quantum
cosmology, not even in perturbative form.

A crucial issue is that of the consistency of higher order terms, such
as those appearing in (\ref{loopFriedmann}), in a context which is no
longer homogeneous. Then, the full anomaly issue strikes and
modifications to the classical constraints are highly restricted: it
is not easy to implement quantum corrections while still maintaining
the same level of general covariance as it is realized classically. If
covariance transformations are broken, the theory will be anomalous
and inconsistent; such transformations could only be deformed by
quantum corrections but must remain present in the same
number. (Effective actions starting from quantum corrected isotropic
equations have been determined
\cite{ActionRhoSquared,ActionLovelock,ActionOrder}. But trying to
embed a finite-dimensional model in a fully inhomogenenous system is
highly ambiguous, and so quantum corrections for inhomogeneities
remain unknown in the presence of higher order corrections such as
(\ref{loopFriedmann}); examples do, however, exist for special modes
\cite{Vector,Tensor} or other effects of loop quantum gravity
\cite{ConstraintAlgebra,ScalarGaugeInv,LTB,SphSymmPSM,LTBII}.)

Consistency issues arise due to general covariance, which implies that
one is dealing with a system of constraints, or an overdetermined set
of equations. While there is only one, trivially consistent constraint
(\ref{loopFriedmann}) in isotropic models, several independent ones
exist when geometries become inhomogeneous. Their algebra under
Poisson brackets obeys certain conditions for the system to be
well-defined, which must also be realized for the quantum
representation. A possibility to sidestep the quantization of
constraints is reduced phase space quantization, where one tries to
find the classical solution space to all constraints and quantizes
it. The usual problems are that (i) constraints may be difficult to
solve completely and (ii) the solution space may be of complicated
structure, for instance in its topological properties, and thus be
difficult to quantize in its own right.

In the context of perturbative inhomogeneities, the first problem does
not arise at least at the linear level since all gauge-invariant
perturbations can easily be written down; see e.g.\
\cite{Bardeen,HamGaugePert,BohmPert,PertObsI,PertObsII,BKdustI,BKdustII}.
For linear perturbations, moreover, topological properties of solution
spaces mostly disappear such that a reduced quantization here may be
viable \cite{BohmPertII,PuchtaMaster}.  Alas, it cannot present a full
theory if it is simply added on to the bouncing background as treated
so far, which was by the Dirac rather than the reduced phase space
procedure. One may deal with the background dynamics also by reduced
phase space techniques \cite{ReducedBounce,ReducedBounceQuant}, but
that would work easily only for a free, massless scalar trivializing
the problem of time. By the Dirac procedure, on the other hand, the
theory can be formulated for general interacting scalars \cite{Blyth},
even though it may be solved easily only in free scalar cases or
perturbations around those.

In a reduced phase space quantization of perturbative inhomogeneities,
no fully defined theory would be available. This may be acceptable if
it can be seen as a valid approximation to some other full theory, but
this is not the case. In fact, in systems not involving the bounce,
where consistent quantizations of perturbative inhomogeneities in loop
quantum gravity are available \cite{ConstraintAlgebra}, one can see
that a reduced phase space quantization would overlook crucial
effects. As shown in \cite{ScalarGaugeInv}, quantum corrections can
induce effective anisotropic stress terms even in systems which
classically have no anisotropic stress. A reduced phase space
formulation based on the classical identities between gauge-invariant
quantities could not see this new quantum effect, and thus must remain
incomplete. Similarly, gauge-fixed treatments (as used e.g.\ in
\cite{HolonomyInfl,BounceCMB} for recent examples) often hide crucial
quantum properties. (Also the more general reduced phase space
formulation of \cite{BKdustI,BKdustII} is subject to these remarks.
Moreover, even in this reduced context, consistency conditions remain
which are yet to be implemented in a possible quantization. While
valuable, these formulations so far do not suffice to see how an
inhomogeneous universe may evolve through a bounce.)

Such a situation makes the task of developing cosmological scenarios
based on bounces difficult. But some indications can nonetheless be
derived from models if they are understood for the local behavior of a
patch of space-time near the moment of its smallest size. How
different patches connect may be impossible to say in the absence of a
fully inhomogeneous description, but the evolution of a single patch
may still carry some surprises. Concrete properties, such as the
density when a patch bounces, may easily change or go away when a
sufficiently general situation is considered. But in addition to such
positive, affirmative properties there are negative ones which tell us
about limitations of what can be said for early stages of the patch.
Negative statements of this form are much more reliable, for if
knowledge of something is constrained in a simple model, it is
unlikely to become better known in a general situation.

There is such a negative property which, rather surprisingly, shows up
even in the exactly solvable model \cite{BeforeBB}. It is not about
classical variables, or the expectation values, but rather about
quantum fluctuations or other moments. As before, we can derive
equations of motion for all the moments, say of second order, forming
a closed set of equations. There are several independent second order
moments and their equations, with correspondingly many initial values
to be chosen for a state. One can cut down the number of parameters by
selecting dynamical coherent states: those that saturate the
uncertainty relations at all times. The calculations are somewhat
lengthy but can be completed \cite{Harmonic}, with the result that
volume fluctuations at early and late times are related by
\begin{eqnarray}
\Delta &:=& \left| \lim_{\phi\to-\infty} \frac{G^{VV}}{\langle\hat{V}\rangle^2}- 
\lim_{\phi\to\infty} \frac{G^{VV}}{\langle\hat{V}\rangle^2}\right| 
\label{asymm}\\
&=& 4\frac{H}{V_{\rm min}}
\sqrt{\left(1-\frac{H^2}{V_{\rm min}^2}+\frac{1}{4}\frac{\hbar^2}{V_{\rm min}^2}\right)
  \frac{(\Delta H)^2}{V_{\rm min}^2} -\frac{1}{4}\frac{\hbar^2}{V_{\rm min}^2}+
  \left(\frac{H^2}{V_{\rm min}^2}-1\right) \frac{(\Delta H)^4}{V_{\rm min}^4}} \nonumber
\end{eqnarray}
where $H$ is the expectation value of the $\phi$-Hamiltonian and $\Delta
H$ its fluctuation. This parameterizes the behavior for all dynamical
coherent states.

\begin{figure}
\begin{center}
\resizebox{0.8\textwidth}{!}{%
  \includegraphics{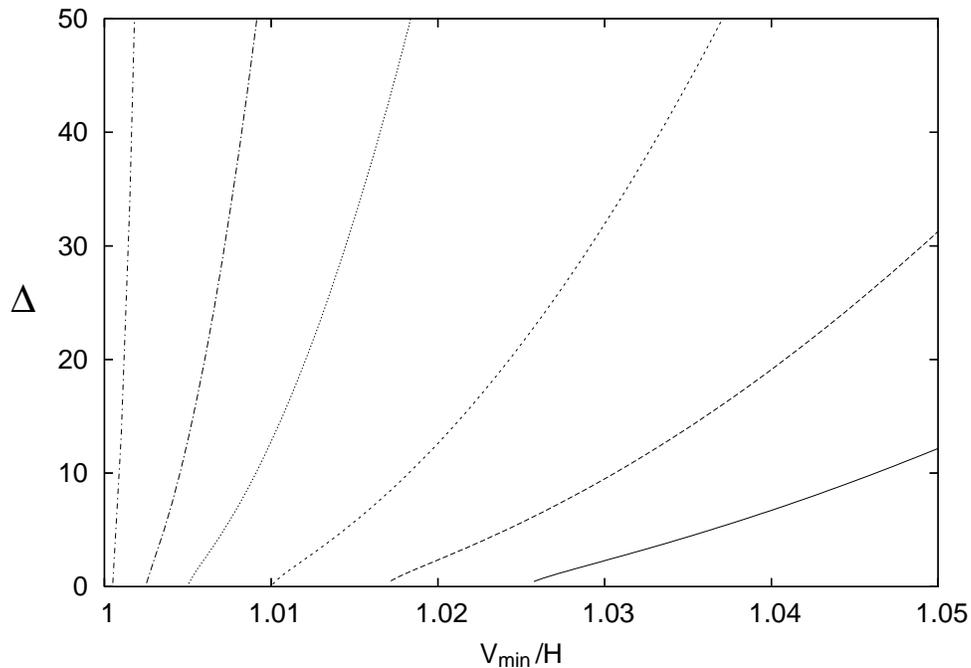}}
\caption{Sensitivity: The asymmetry $\Delta$ of volume
  fluctuations from  (\ref{asymm}), depending on the ratio $V_{\rm min}/H$ with
  $H$ the value of the scalar momentum as a measure
  for the amount of matter. Different curves correspond to
  various values of $H$, growing to the left. Thus, the steep leftmost
  curves are obtained for a universe with a large amount of matter,
  the more realistic scenario within the simple solvable
  models used here. The asymmetry (\ref{asymm}) depends very
  sensitively on the initial values that determine $V_{\rm min}/H$;
  unrealistically sensitive measurements would be required at one side
  of the bounce to determine the volume fluctuations at the other
  side. It is practically impossible to recover the complete state due
  to this cosmic forgetfulness \cite{Harmonic}.
  \label{f:Sens}}
\end{center}
\end{figure}

Of particular interest is the behavior when $H$ is large, which means
that one would use the model for a patch containing a large amount of
matter. As shown in Fig.~\ref{f:Sens}, the asymmetry in such a case
depends very sensitively on the parameters, for instance the ratio
$V_{\rm min}/H$. Moreover, its value can differ significantly from
one; fluctuations of the state by no means have to remain unchanged
when phases before and after the bounce are considered.  There is a
degree of cosmic forgetfulness \cite{BeforeBB}: due to the high
sensitivity it is practically impossible to recover the full state
before the bounce from its properties after the bounce.

\section{An arrow of moments}

In our discussion of the covariance, the big bang, resolved to a
bounce, did not appear special in any way regarding the direction of
time. It did not suggest a turn-around in the rate of change of the
covariance. Had it done so, it would have led us to conclude that
$G^{VP}$ cannot serve as a good time in that phase, rather than
suggesting a flip of the arrow of time.

The role of moments concerning the arrow of time is more subtle. We
will first reformulate the usual context to see how it may be related
to evolving quantum states. One often says that what distinguishes the
past from the future is that we remember the former and try to predict
the latter. It may be more honest to define the past as what we can
(and typically do) forget.  For human behavior, one of the most
important and most annoying consequences of the arrow of time is
indeed forgetfulness. In a more general sense, this is true also for
thermodynamical systems, although it may not be so clear whether this
is really annoying. A thermodynamical system evolving toward
equilibrium forgets any sense of being special as it might have been
encoded in its initial configuration. In quantum cosmology, even the
whole universe has a case of cosmic forgetfulness which one may relate
to the arrow of time.\footnote{In thermodynamics, coarse-graining
  plays an important role. Cosmic forgetfulness may be interpreted as
  forcing us to coarse-grain over many of the quantum variables. One
  should also note that cosmic forgetfulness is much stronger than
  decoherence (see e.g.\ \cite{OpenArrow}) since it appears even in
  exactly solvable models. It takes into account the specific dynamics
  of loop quantum cosmology, rather than being a generic property of
  quantum systems with many degrees of freedom.}

To illustrate this, we return to the resummed solvable model of loop
quantum gravity. Now considering its own moments for $V$ and $J$, we
can look for all choices giving rise to dynamical coherent states:
evolving states which saturate the uncertainty relations at all times.
Such states provide the best control one may have on a quantum system,
and thus highlight when anything becomes inaccessible --- for instance
by being forgotten. As already described, this is exactly what
happens.  Although one could not easily use the solvable model to draw
strong conclusions about the universe before the big bang, what it
tells us about limitations has to be taken seriously.\footnote{Cosmic
  forgetfulness has been perceived as a challenge, heroically taken up
  in \cite{BounceRecall} by deriving bounds alternative to
  (\ref{asymm}) for semiclassical states. However, those bounds are
  much weaker, allowing changes in the fluctuations by several orders
  of magnitude \cite{RecallComment}. (Also this renewed challenge has
  been taken up in \cite{BounceRecallReply}, though less heroically so.)}

The kind of cosmic forgetfulness realized in this model provides an
orientation of time, telling us not only which of the properties
before the big bang can be forgotten, but also what direction ``before
the big bang'' is. An observer after the bounce would be unable to
reconstruct the full state before the bounce, but could easily predict
the future development toward larger volume. This arrow agrees with
the standard notion.

Now asking how an observer before the big bang would experience the
same situation, the answer is also clear: such an observer would be
unable to determine the precise state at larger values of $\phi$
beyond the bounce, but could easily extrapolate the state to smaller
values of $\phi$. The state at smaller values of $\phi$ can be
predicted, while the state at large values of $\phi$ is forgotten once
the bounce is penetrated. Since one cannot forget the future, such an
observer must be attributed a reversed arrow of time, pointing toward
smaller $\phi$. At the bounce, two arrows would emerge pointing in
opposite directions as far as $\phi$ is concerned. In this sense, the
model resembles
\cite{KieferZeh,SteadyStateInflation,SteadyStateInflationII,CarrollChen}.

While degrees of freedom propagating in a bouncing universe still have
to be understood much better, indications do exist that what appears
as a simple bounce in homogeneous models may have to be interpreted
rather differently when degrees of freedom other than the purely
classical homogeneous ones are considered. Here, this has been
discussed for homogeneous quantum degrees of freedom; inhomogeneities
will be the next crucial and decisive step.\footnote{Numerical
indications for a similarly sensitive behavior of inhomogeneities
\cite{InhomThroughBounce} already exist from Gowdy models with a
loop-quantized homogeneous background \cite{Hybrid}.}

\section{Conclusions}

``If he had smiled why would he have smiled? To reflect that each one
who enters imagines himself to be the first to enter whereas he is
always the last term of a preceding series even if the first term of a
succeeding one, each imagining himself to be first, last, only and
alone, whereas he is neither first nor last nor only nor alone in a
series originating in and repeated to infinity.''\footnote{James
  Joyce: Ulysses} This describes the
thoughts of Leonard Bloom after a long eventful day. Will we be led to
similar thoughts after a long eventful journey in quantum gravity?

If we cannot reconstruct the entire past, we may as well forget about
it. The part of the universe we see would appear to have originated
with its big bang, even though a theoretical formulation, but only the
theoretical formulation, may contain a pre-history. Two questions
should immediately be asked: Would this be testable? And why would we
not apply Occam's razor to the pre-history?  We could clearly not
directly test whether there is a part of the history of the universe
that is inaccessible. But we may attempt to access it and, if we
succeed, falsify the claim; this makes it scientifically viable as a
hypothesis. More importantly, the underlying scenario would have
further implications for the structures we see after the big
bang. Then, we would have an option to test such a model indirectly.

Why do we then consider the pre-history as part of the mathematical
modelling? Also this has its justification. Describing a true physical
beginning of the universe, where nothing would turn into something,
has proved to be challenging.  Pretending that there was something
before the big bang and describing it by deterministic but forgetful
equations may be the best solution to deal with a beginningless
beginning, even though we may not be able to use those equations to
fully reconstruct the past.

Taking the simplest models of loop quantum cosmology at face value is
often seen as suggesting the big bang transition to be viewed as a
smooth bounce, as one further element not just in a long history of
the universe itself but also in a long history of bouncing
cosmological models \cite{BounceReview}. Some indications, however,
suggest otherwise. The bloomy scenario of loop quantum cosmology may
well be this: a universe whose time-reversed pre-history we cannot
access but which we grasp in the form of initial conditions it
provides for our accessible part; a pseudo-beginning
\cite{TimeBeforeTime}; an orphan universe, shown the rear-end by
whatever preceded (and possibly created) it.

\bigskip

\noindent {\bf Acknowledgements:} 

\medskip

This work was partially supported by NSF grant PHY0748336 and a grant
from the Foundational Questions Institute (FQXi). The author is
grateful to R\"udiger Vaas for an invitation to contribute to the
volume ``The Arrow of Time'' and for several suggestions on the
manuscript.

\section*{Appendix: A momentous formulation of quantum mechanics}

Quantum dynamics can usefully be described in terms of the moments
(\ref{Moments}) of a state.  Taken together, they form an
infinite-dimensional phase space which can be used to describe the
quantum system. At order $a+b=2$ we have the fluctuations
$G^{2,0}=(\Delta V)^2$ and $G^{0,2}=(\Delta P)^2$ as well as the
covariance $G^{1,1}=
\frac{1}{2}\langle\hat{V}\hat{P}+\hat{P}\hat{V}\rangle-
\langle\hat{V}\rangle\langle\hat{P}\rangle$.  While independent
variables, the moments cannot be chosen arbitrarily. They are subject
to constraints, most importantly the uncertainty relation
\[
 G^{VV}G^{PP}-(G^{VP})^2\geq \frac{\hbar^2}{4}\,.
\]
Poisson brackets between the moments can be computed using the general
identity
\begin{equation}
  \{\langle\hat{A}\rangle,\langle\hat{B}\rangle\}= 
\frac{\langle[\hat{A},\hat{B}]\rangle}{i\hbar}
\end{equation}
as well as linearity and the Leibniz rule. This immediately gives
$\{\langle\hat{V}\rangle,\langle\hat{P}\rangle\}=1$ and, e.g.,
$\{G^{VV},G^{PP}\}= 4G^{VP}$. (See \cite{EffAc,Karpacz} for further details.)

The moments allow a convenient description of quantum evolution
without having to take the usual detour of solving for states first,
followed by computing expectation values. Instead, expectation values
obey the general evolution law
\[
 \frac{\md \langle\hat{O}\rangle}{\md\phi}= 
\frac{\langle[\hat{O},\hat{H}]\rangle}{i\hbar}
\]
which can be used to derive a coupled set of equations of motion for
expectation values together with the moments. For non-polynomial
$\hat{H}$, it may be difficult to compute the commutator, followed by
taking an expectation value. Semiclassical equations can more easily
be obtained in an expansion by moments, which is formally analogous to
background-field expansion around expectation values. We write
\cite{EffAc}
\begin{eqnarray}
\langle H(\hat{V},\hat{P})\rangle&=&\langle
H(\langle\hat{V}\rangle+(\hat{V}-\langle\hat{V}\rangle),
\langle\hat{P}\rangle+(\hat{P}-\langle\hat{P}\rangle))\rangle\\
&=& H(\langle\hat{V}\rangle,\langle\hat{P}\rangle)+\sum_{a,b:a+b\geq 2} 
\frac{1}{a!b!}
\frac{\partial^{a+b}H(\langle\hat{V}\rangle,\langle\hat{P}\rangle)}{\partial \langle\hat{V}\rangle^a\partial \langle\hat{P}\rangle^b}G^{a,b}
\end{eqnarray}
and use this in
\begin{equation}
 \frac{\langle[\hat{O},\hat{H}]\rangle}{i\hbar}= \{
 \langle\hat{O}\rangle, \langle H(\hat{V},\hat{P})\rangle\}\,.
\end{equation}
Poisson relations between the moments then provide all equations of
motion order by order in the moments.

For the cosmological
systems with Hamiltonian (\ref{pphi}) introduced before, we have
\begin{eqnarray*}
 \frac{\md \langle\hat{V}\rangle}{\md\phi} &=& \frac{3}{2}
\frac{\langle\hat{V}\rangle\langle\hat{P}\rangle}{\sqrt{\langle\hat{P}\rangle^2+|\Lambda|}}
-\frac{9}{4}|\Lambda| \frac{\langle\hat{V}\rangle\langle\hat{P}\rangle}{(\langle\hat{P}\rangle^2+|\Lambda|)^{5/2}} G^{PP}
+\frac{3}{2}|\Lambda| \frac{G^{VP}}{(\langle\hat{P}\rangle^2+|\Lambda|)^{3/2}}+\cdots \\ 
\frac{\md \langle\hat{P}\rangle}{\md\phi} &=& -\frac{3}{2}\sqrt{\langle\hat{P}\rangle^2+|\Lambda|}- \frac{3}{4}|\Lambda|
\frac{G^{PP}}{(\langle\hat{P}\rangle^2+|\Lambda|)^{3/2}}+\cdots
\end{eqnarray*}
expanded by the moments (kept here to second order only).
This is accompanied by the evolution of moments
\begin{eqnarray}
 \frac{\md G^{PP}}{\md\phi} &=& -3\frac{\langle\hat{P}\rangle}{\sqrt{\langle\hat{P}\rangle^2+|\Lambda|}} G^{PP}+\cdots \label{GPP} \\
 \frac{\md G^{VP}}{\md\phi} &=& \frac{3}{2}|\Lambda|
\frac{\langle\hat{V}\rangle}{(\langle\hat{P}\rangle^2+|\Lambda|)^{3/2}}G^{PP}+\cdots \label{GVP}\\
 \frac{\md G^{VV}}{\md\phi} &=& 3 |\Lambda|
\frac{\langle\hat{V}\rangle}{(\langle\hat{P}\rangle^2+|\Lambda|)^{3/2}} G^{VP}
 +3\frac{\langle\hat{P}\rangle}{\sqrt{\langle\hat{P}\rangle^2+|\Lambda|}} G^{VV}+\cdots\,. \label{GVV}
\end{eqnarray}
Solving or analyzing this coupled set of equations would tell us how
the state changes its shape by the evolving moments, and how this
back-reacts on the motion of expectation values. In some regimes it is
possible to summarize the effect of all moments in an effective
potential for expectation values depending only on the classical
variables. But in general, higher-dimensional effective systems
including the moments as independent variables are required.

\end{document}